\documentstyle[twoside,fleqn,espcrc2,epsfig]{article}
 
\newcommand{\be}{\begin{equation}}
\newcommand{\ee}{\end{equation}}
\newcommand{\bea}{\begin{eqnarray}}
\newcommand{\eea}{\end{eqnarray}}
\newcommand{\tp}{\tilde \pi}
\newcommand{\ts}{\tilde \sigma}
\newcommand{\tchi}{\tilde \chi}
\newcommand{\nn}{\nonumber}

\title{Quenched Spectroscopy for the $N=1$ Super-Yang--Mills Theory}
\author{A.~Donini\address{I.N.F.N. and Dipartimento di Fisica 
dell' Universit\'a di Roma ``La Sapienza'' \\ Piazza Aldo Moro 2, I-00185 
Roma, Italy.} \thanks{Talk presented by A. Donini.}, 
M.~Guagnelli\address{Theory Division, CERN, 1211 Geneva 23, 
Switzerland.}, P.~Hernandez$^{\rm b}$, 
A.~Vladikas\address{I.N.F.N. and Dipartimento di Fisica 
dell' Universit\'a di Roma ``Tor Vergata'' \\ Via della Ricerca Scientifica 1, 
I-00133 Rome, Italy.}}

\begin{document}
\begin{abstract} 
We present results for the Quenched SU(2) $N=1$ Super-Yang--Mills 
spectrum at $\beta=2.6$, on a $V=16^3 \times 32$ lattice ,
in the OZI approximation. This is a first step towards the understanding of 
the chiral limit of lattice $N=1$ SUSY. 
\end{abstract}
 
\maketitle

\section{Introduction}
Softly broken $N=1$ supersymmetry gives a natural solution to the 
long-standing hierarchy problem arising from the Standard Model, considered 
as a low-energy effective model of a more fundamental theory which includes 
gravity. 
Moreover, analytical solutions have been found in a wide class of 
supersymmetric gauge theories \cite{Sewi}. 
The most remarkable results have been obtained in the context of 
extended supersymmetry ($N > 1$), where the renormalization properties are 
severely constrained. Thus, the study of supersymmetric gauge theories on 
the lattice is of great importance for the confirmation of the predicted 
non-perturbative behaviour for $N > 1$ and the extension of our
understanding to the phenomenologically relevant softly broken $N=1$ case.

The simplest non-abelian supersymmetric gauge theory is $N=1$ SU(2) 
Super-Yang--Mills (SYM): this theory contains gluons and gluinos (Majorana
fermions in the adjoint representation) and no matter fields. 
%The lattice formulation of the $N=1$ SYM theory has been given in 
%ref.\cite{cv}. The gauge action $S_g$ is the usual sum over plaquettes. 
%Majorana gluino fields are implemented as in QCD, by adding a Wilson term. 
%The fermionic action is
%\be
%S_f = \bar \lambda^a(x)
%\Delta^{ab}(x,y) \lambda^b(y) ,
%\ee
%where $a,b$ are color indices of the adjoint representation 
%($a,b = 1, \dots, N_c^2 - 1$). The fermion matrix $\Delta$ describe 
%Majorana fermions with $\Delta = C M$, where $C$ is the charge conjugation 
%matrix and $M$ is antisymmetric \cite{m2}. 
A lattice formulation of the $N=1$ SYM theory was given in ref.\cite{cv}. 
The path integral is 
\be
Z = \int DU e^{-S_g} det(\Delta)^{1/2} e^{-\bar{\eta^a} [\Delta^{-1}]_{ab} 
\eta^b}
\ee
where $S_g$ is the usual sum over plaquettes; $\Delta_{ab}$ is the fermionic 
matrix (containing a Wilson term) and $a,b = 1, \dots, N_c^2 - 1$ are color
indices.
The Majorana nature of the fermion fields is 
reflected in the power $1/2$ of the fermion determinant \cite{m2}.
Some preliminary unquenched results for gauge observables 
have been presented in refs.\cite{m2,dg}.

Supersymmetry imposes that the gluon and the gluino should be degenerate
in mass (hence, due to gauge symmetry, both should be massless).
However, both chiral symmetry and supersymmetry are explicitly broken on the
lattice by the Wilson term, which gives mass to the gluino.
Moreover, supersymmetry is also broken by the space-time discretization.
In order to recover the desired continuum theory, both symmetries should be 
restored. In ref.\cite{cv} it has been shown that the chiral and the
supersymmetric Ward Identities (WI) are simultaneously satisfied in the
continuum limit, by tuning the bare gluino mass $m_0$ to the critical value 
$m_c$ for which the renormalized gluino mass vanishes.

As in QCD, due to confinement, the low-energy spectrum is expected to
consist of colorless composite fields of gluons and gluinos, 
which can be described by a low-energy effective 
supersymmetric lagrangian. Due to supersymmetry, the gluino-gluon bound 
belong to supermultiplets. The lowest-lying chiral
supermultiplet is \cite{vy}
\be
\Phi = ( \tp , \ts , \tchi , F , G ) ,
\ee
where $\tp = \bar \lambda^a \gamma_5 \lambda^a$ is a pseudoscalar,
$\ts = \bar \lambda^a \lambda^a$ is a scalar and
$\tchi = \frac{1}{2} \sigma_{\mu\nu} F^a_{\mu\nu} \lambda^a$ is a fermion.
The fields $F$ and $G$ (scalar and pseudoscalar respectively) are
auxiliary fields with no dynamics. In the supersymmetric limit, these
particles should be degenerate in mass \cite{vy}. 

The composite $\tp$ pseudoscalar is analogous to the QCD singlet
$\eta^\prime$, since a single flavour of gluinos is present in the theory.
Its two-point correlation function $C_{\tp}(t)$ contains connected 
as well as disconnected diagrams. In \cite{vy} it has been suggested that,
in the OZI approximation (where only the connected part of $C_{\tp}(t)$
is retained), $\tp$ should behave like the QCD pion. This means that in this
approximation $\tp$ should become massless at some critical value $m_c$
of the bare gluino mass, whereas $\ts$ and $\tchi$ should remain massive. 
In the full theory, the disconnected part of $C_{\tp}(t)$ should give mass 
to $\tp$ and restore the predicted mass degeneracy between all the particles 
belonging to the $\Phi$ supermultiplet (since chiral symmetry and
supersymmetry are simultaneously restored at $m_c$ \cite{cv}).

\section{Quenched Super-Yang-Mills theory}

Since supersymmetry requires an exact balance of the bosonic and fermionic
degrees of freedom, the quenched approximation is not necessarily a good
approximation to a supersymmetric theory. Nevertheless, it is useful
to study the quenched spectrum as a preliminary step towards the true SYM.

We present results for the supermultiplet $\Phi$ for $\tp,\ts$ and $\tchi$, 
obtained at $\beta=2.6$ on a $V=16^3 \times 32$ lattice
in the OZI approximation. All statistical errors have been computed
with the jacknife method. 
Some preliminary quenched results for $\tchi$ have been presented in 
\cite{Montlatt} (see also \cite{Montlatt2}). 

We obtain the pseudoscalar and scalar masses from a sample of 950 gluino
propagators at four different values of the Wilson hopping parameter:
$K = 0.174, 0.178, 0.182, 0.184$. These values have been chosen in 
order to simulate relatively light gluinos. In the case of the $\tp$
two-point correlation function, the effective mass has not settled fully to a
plateau; the lattice is somewhat short in the time-direction. We extract a 
reliable $\tp$ mass by performing two-mass fits of the correlation function. 
In the case of the $\ts$ correlation function a plateau is observed within 
the statistical errors (much bigger than for $\tp$). Hence, we can extract 
the $\ts$ mass from a one-mass fit. We stress that, in contrast to the 
measurements of the QCD scalar, no smearing has been required.
The results for the $\tp$ and $\ts$ masses are 
presented in fig.\ref{fig2}. The $\tp$ mass shows the expected dependence 
on the Wilson hopping parameter: namely, $m^2_{\tp}$ decreases linearly
with $K$. The critical value where the gluino becomes massless is
\be
K_c = 0.18752(9)
\label{eq:kc_1}
\ee
(with one-mass fits we obtain $K_c = 0.18759(2)$ ).
%pippo
%___________________________________________________________________________
\begin{figure}[t]   % produce figure here
\vspace{9pt}
\centerline{\epsfig{figure=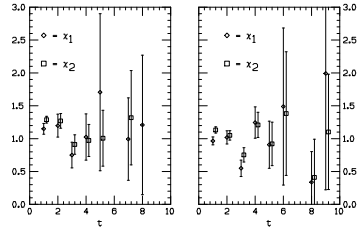,height=5cm,angle=90}}
\caption{Effective masses for $\tchi$ with $n_s = 4$ (left)
and $n_s = 8$ (right) as a function of time at $K=0.182$. Diamonds
refer to $C_1(t)$ and squares to $C_2(t)$.} 
\label{fig1}
\end{figure}
%___________________________________________________________________________

A numerically independent measure of $K_c$ can be extracted from the
quenched OZI chiral Ward Identity. In a theory with two flavours of gluinos,
the non-singlet axial current $A^{ns}_\mu = \bar \lambda^a \gamma_\mu 
\gamma_5 \lambda^a$ satisfies the usual PCAC relation:
\be
Z_A \partial_\mu A^{ns}_\mu = 2 (m_0 - \bar m ) P^{ns},
\ee
where $P^{ns} = \bar \lambda^a \gamma_5 \lambda^a$ is the non-singlet
pseudoscalar density, $Z_A$ the finite renormalization constant of
$A_\mu^{ns}$ and $m_0 - \bar m = 0 $ at $K_c$. In the quenched 
approximation, the following identities hold:
\bea
\langle \partial_\mu A^s_\mu P^s \rangle \vert_{OZI} &=& 
\langle \partial_\mu A^{ns}_\mu P^{ns} \rangle , \\
\langle P^s P^s \rangle \vert_{OZI} &=& 
\langle P^{ns} P^{ns} \rangle \nn ,
\eea
where $A^s_\mu,P^s$ are the singlet axial current and pseudoscalar
density respectively. Hence, 
\be
2 \rho = \frac{\langle \partial_\mu A^s_\mu P^s \rangle}{\langle 
P^s P^s \rangle} \Bigg \vert _{OZI} = \frac{2(m_0 - \bar m)}{Z_A} .
\label{eq:2rho}
\ee
The quantity $2\rho$ should vanish in the chiral limit. We obtain
in this way
\be
K_c = 0.18762(4)
\ee
in perfect agreement with eq.(\ref{eq:kc_1}). 
For more details, see \cite{dghv}.

Using its properties under the discrete symmetries, it is easy to show 
\cite{dghv} that the two-point correlation function of $\tchi$ is described 
by just two form factors:
\be
C_{\tchi}(t) = C_1(t) I + C_2(t) \gamma_0.
\ee
Moreover, $C_{\tchi}(t)$ is OZI-blind. In order to reduce contact terms
in the time-direction (note that $\tchi$ is an extended object, 
being proportional to $\sigma_{\mu\nu} F^a_{\mu\nu}$) we use the spatial
operator $\tchi^s = \frac{1}{2} \sigma_{ij} F^a_{ij} \lambda^a$, whose
correlation function should behave as $C_{\tchi}(t)$ at large times. We 
implemented the smearing 
of ref.\cite{APE_sme} in $F^a_{ij}$ by replacing the link $U_\mu(x)$ with
$U_\mu(x)$ plus the staple multiplied by a tuning parameter $\epsilon$ (in our
case, $\epsilon$ has been fixed at 0.5). This smearing has been repeated 
$n_s=4$ and 8 times.
%pippo
In fig.\ref{fig1} we present results obtained for the effective mass of
$\tchi$ with $n_s = 4$ (500 gluino propagators) and $n_s = 8$ (450 propagators)
at a fixed $K$. We show results from both $C_1(t)$ and $C_2(t)$. 
A short plateau appears at very small times.
We extract the $\tchi$ effective mass from the $C_2(t)$ form factor at 
$n_s = 4$. No signal at all was observed without smearing.

In fig.\ref{fig2} we present the masses of $\ts$ and $\tchi$, the squared
$\tp$ mass and the $2\rho$ parameter. Linear extrapolations at $K_c$
are also shown.
%___________________________________________________________________________
\begin{figure}[t]   % produce figure here
\vspace{9pt}
\centerline{\epsfig{figure=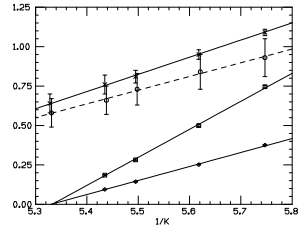,height=5cm,angle=90}}
\caption{Results for the $\ts$ mass (crosses), the $\tchi$ mass (circles),
the $\tp$ squared mass (squares) and the $2\rho$ parameter (diamonds) as a
function of $1/K$.}
\label{fig2}
\end{figure}
%___________________________________________________________________________
It can be seen that the $\ts$ and $\tchi$ masses at $K_c$ are similar
in magnitude. This could be a first signal of the recovery of 
supersymmetry at $K_c$, but must be carefully verified in the context of a 
full unquenched non-OZI simulation. 

%\section{Conclusions}
%Supersymmetric gauge theories can be studied on the lattice with the 
%present computing resources. In the quenched approximation, the OZI
%pseudoscalar $\tp$ does behave as suggested in \cite{vy}. The critical 
%value of the hopping parameter $K_c$ extracted in this way is perfectly
%compatible with that obtained by using the quenched OZI $2\rho$ parameter.
%Once a neat method to recover
%the supersymmetric limit is achieved, the non-perturbative properties of 
%$N=1$ supersymmetric gauge theories could be thoroughtfully studied.

%\section*{Acknowledgements.}

We thank I. Montvay, G. C. Rossi, M. Testa and G. Veneziano for useful
discussions. We also thank the Rome-1 APE group, and in particular G.
Martinelli, for providing us with the necessary computing resources.

\end{document}